\documentclass[12pt]{iopart}

\usepackage{iopams}
\usepackage{epsfig}
\begin{document}

\title{ Quarkonium production in coherent $pp/AA$  collisions and small-$x$ physics}

\author{V.P. Gon\c{c}alves and M.V.T. Machado}

\address{Universidade Federal de
Pelotas
\\
Caixa Postal 354, CEP 96010-090, Pelotas, RS, Brazil}

\begin{abstract}
In this contribution we study the photoproduction of quarkonium in
coherent  proton-proton and nucleus-nucleus  interactions at the LHC.
The integrated cross sections and rapidity distributions are
estimated using the Color Glass Condensate (CGC) formalism, which
takes into account the parton saturation effects at high energies. Nuclear shadowing effects are also taken into account.

\end{abstract}


In this contribution we study the   photoproduction of  vector
mesons in the coherent $pp/AA$ interactions at the LHC energies. The
main advantage of using colliding hadrons and nuclear beams for
studying photon induced interactions is the high equivalent photon
energies and luminosities that can be achieved at existing and
future accelerators (For a review see Ref. \cite{Goncalves:2005sn}).
 Consequently, studies of $\gamma p$ interactions
at LHC could provide valuable information on the QCD dynamics at
high energies. The basic idea in coherent  hadron collisions is that
the total cross section for a given process can be factorized in
terms of the equivalent flux of photons of the hadron projectile and
the photon-photon or photon-target production cross section. In
exclusive processes, a certain particle is produced while the target
remains in the ground state (or is only internally excited). The
typical examples of these processes are the exclusive vector meson
production, described by the process $\gamma h \rightarrow V h$ ($V
= \rho, J/\Psi, \Upsilon$). In the last years we have discussed this
 process in detail considering $pp$ \cite{per4},
$pA$ \cite{perpa} and $AA$ \cite{per4} collisions as an alternative
to investigate the QCD dynamics at high energies. Here, we revised
these results and present for the first time our predictions for the
$\Upsilon$ production.

The cross section for the  photoproduction of a vector meson $X$ in
an ultra-peripheral hadron-hadron collision is given by
\begin{eqnarray}
\sigma (h_1 h_2 \rightarrow h_1 h_2 X)\, = \int
\limits_{\omega_{min}}^{\infty} d\omega\,
\frac{dN_{\gamma}(\omega)}{d\omega}\,\sigma_{\gamma h \rightarrow X
h} \left(W_{\gamma h}^2\right)\,, \label{sigAA}
\end{eqnarray}
where $\omega$ is the photon energy   and $
\frac{dN_{\gamma}(\omega)}{d\omega}$ is the equivalent flux of
photons from a charged hadron. The total cross section for vector
meson photoproduction is  calculated considering the color dipole
approach, which is directly related with the dipole-target forward
amplitude ${\cal{N}}$. In the Color Glass Condensate (CGC) formalism
 (See e.g. \cite{hdqcd}),  ${\cal{N}}$ encodes all the information
about the hadronic scattering, and thus about the non-linear and
quantum effects in the hadron wave function. In our analyzes we have
used the phenomenological saturation model proposed in Refs.
\cite{gbw,iim}. Nuclear effects are also properly taken into account.

\begin{figure}
\begin{tabular}{cc}
\psfig{file=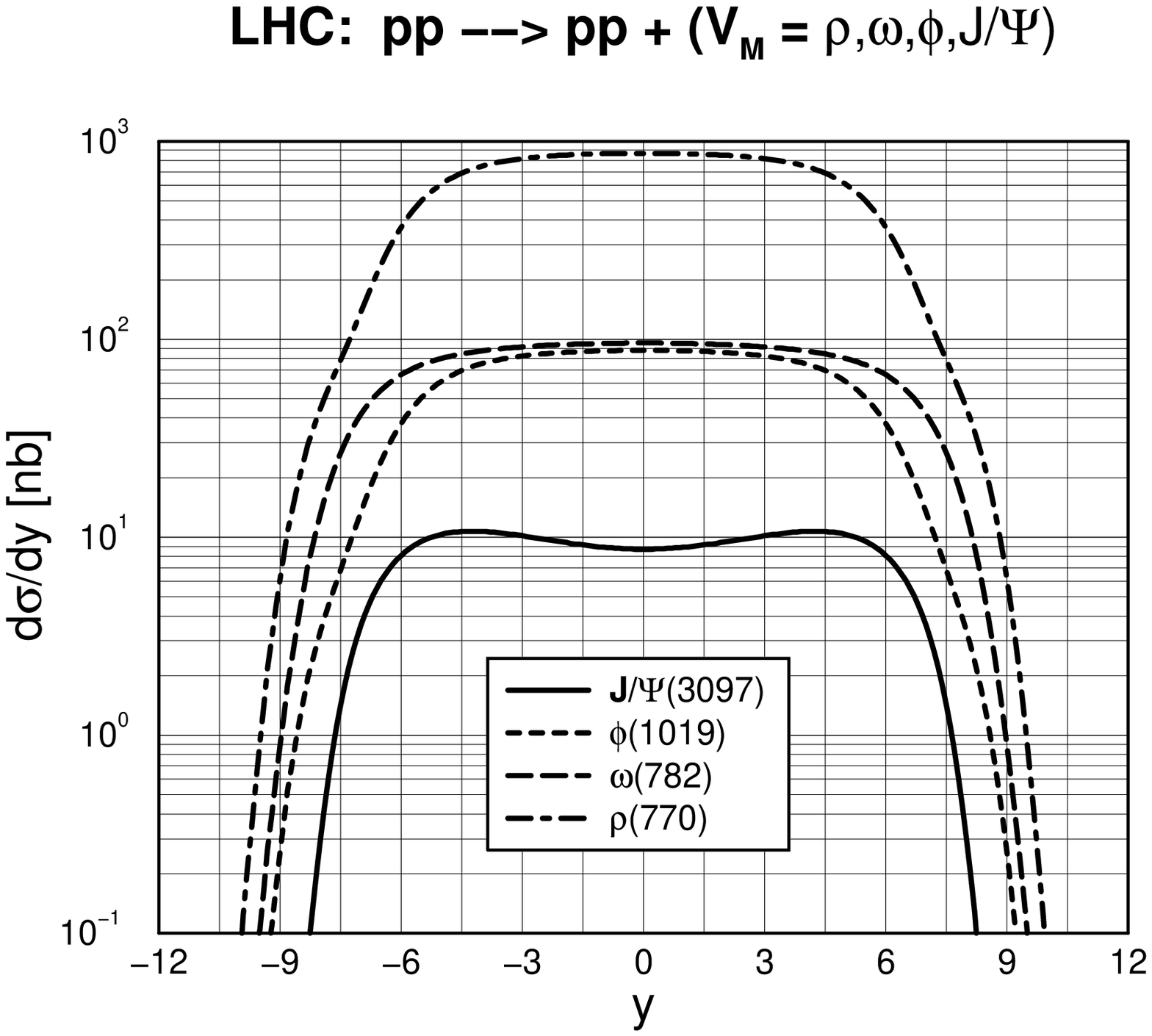,width=80mm} &
\psfig{file=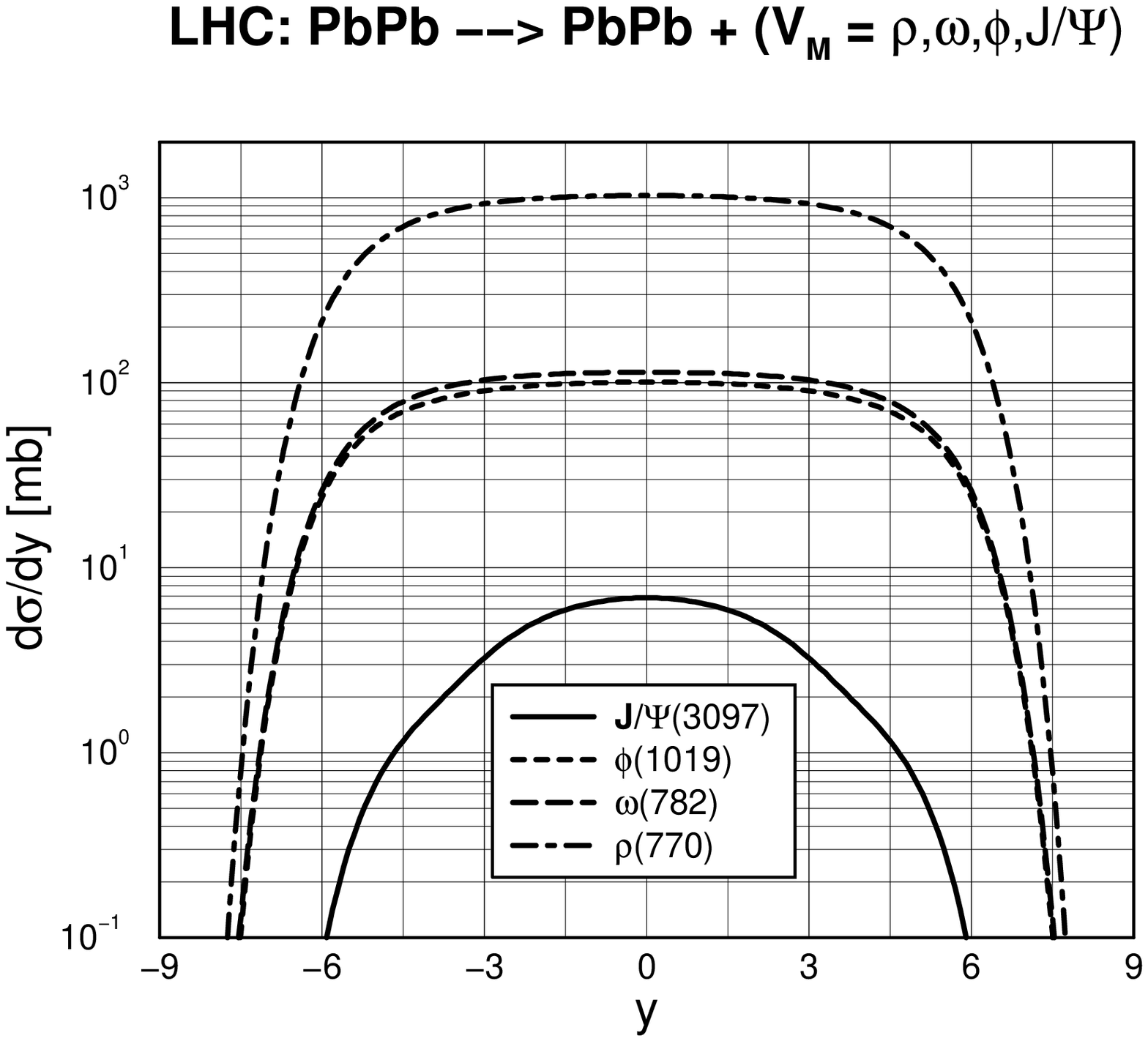,width=80mm}
\end{tabular}
 \caption{\it The rapidity distribution for nuclear vector meson  photoproduction on coherent $pp$ (left panel) and $AA$ (right panel)  reactions at the LHC.}
\label{fig1}
\end{figure}

Our predictions for the rapidity distributions are presented in Fig.
\ref{fig1} and for the total cross section in Table \ref{tab1}. The main uncertainties are the photon flux, the quark mass and the size of nuclear effects for the photonuclear case. In addition, specific predictions for $\rho$ and $J/\Psi$ phoproduction in $pA$ collisions can be found in Ref. \cite{perpa}. The
rates are very high, mostly for light mesons. Although the rates are
lower than hadroproduction, the coherent photoproduction signal
would be clearly separated by applying a transverse momentum cut
$p_T<1$ and two rapidity gaps in the final state.

\begin{table}
\begin{center}
\begin{tabular} {||c|c|c|c|c|c||}
\hline \hline
    & $\Upsilon\,(9460)$ & $J/\Psi\,(3097)$ & $\phi\,(1019)$ & $\omega\,(782)$ & $\rho\,(770)$  \\
\hline \hline
 pp & 0.8 nb  & 132 nb &  980 nb & 1.24 $\mu$b & 9.75 $\mu$b \\
\hline
 CaCa & 9.7 $\mu$b &436 $\mu$b & 12 mb & 14 mb &  128 mb \\
\hline
  PbPb & 96 $\mu$b & 41.5 mb &  998 mb & 1131 mb & 10069 mb \\
\hline \hline
\end{tabular}
\end{center}
\caption{\it The integrated cross section for nuclear vector mesons
photoproduction in coherent $pp$ and $AA$ collisions at the LHC.}
\label{tab1}
\end{table}



\section*{References}
\begin{thebibliography}{10}

\bibitem{Goncalves:2005sn}
  V.~P.~Goncalves and M.~V.~T.~Machado,
  J.\ Phys.\ G {\bf 32} (2006) 295
  [arXiv:hep-ph/0506331].

\bibitem{per4}
  V.~P.~Goncalves and M.~V.~T.~Machado,
  Eur.\ Phys.\ J.\  C {\bf 40} (2005) 519
  [arXiv:hep-ph/0501099].

\bibitem{perpa}
  V.~P.~Goncalves and M.~V.~T.~Machado,
  Phys.\ Rev.\  C {\bf 73} (2006) 044902
  [arXiv:hep-ph/0511183].


\bibitem{hdqcd}
J.~Jalilian-Marian and Y.~V.~Kovchegov, Prog.\ Part.\ Nucl.\ Phys.\
{\bf 56} (2006) 104.

\bibitem{gbw} K. Golec-Biernat and  M. W\"usthoff,  Phys. Rev. D {\bf 59}  014017 (1999) 014017.

\bibitem{iim}
  E.~Iancu, K.~Itakura and S.~Munier,
  Phys.\ Lett.\ B {\bf 590} (2004) 199.
\end{thebibliography}
\end{document}